\documentstyle[12pt]{article}

\begin{document}
\def\ba{\begin{array}}
\def\ea{\end{array}}
\def\bc{\begin{center}}
\def\ec{\end{center}}
\def\be{\begin{eqnarray*}}
\def\ee{\end{eqnarray*}}
\def\bi{\begin{itemize}}
\def\ei{\end{itemize}}
\def\bm{\left(\ba}
\def\em{\ea\right)}
\def\bn{\begin{eqnarray}}
\def\en{\end{eqnarray}}
\def\bt{\begin{tabular}}
\def\et{\end{tabular}}
\def\bdes{\begin{description}}
\def\edes{\end{description}}

\def\ds{\displaystyle}
\def\sss{\scriptscriptstyle}
\def\mc{\multicolumn}
\def\nc{\nonumber \\}

\def\a{\alpha}
\def\b{\beta}
\def\d{\delta}
\def\e{\epsilon}
\def\f{\frac}
\def\g{\gamma}
\def\h{\hbar}
\def\k{\kappa}
\def\l{\lambda}
\def\n{\nabla}
\def\o{\omega}
\def\p{\partial}
\def\s{\sigma}
\def\t{\times}
\def\v{\varepsilon}
\def\D{\Delta}
\def\G{\Gamma}
\def\L{\Lambda}
\def\Si{\Sigma}
\def\O{\Omega}

\def\and{\mbox{\ and\ }}
\def\at{\mbox{at\ }}
\def\because{\mbox{because\ }}
\def\det{\mbox{det\ }}
\def\for{\mbox{\ for\ }}
\def\in{\mbox{in\ }}
\def\tr{\mbox{tr\ }}
\def\on{\mbox{on\ }}
\def\out{\mbox{out\ }}
\def\otherwise{\mbox{otherwise\ }}
\def\then{\mbox{then\ }}
\def\with{\mbox{with\ }}

\def\ab{{\a\b}}
\def\ij{{ij}}
\def\ji{{ji}}
\def\jk{{jk}}
\def\ik{{ik}}
\def\ls{{\l\s}}
\def\mn{{\mu\nu}}
\def\mnl{{\mu\nu\l}}
\def\mnr{{\mu\nu\rho}}

\def\Lmn{{\L^\mu_\nu}}
\def\lmn{{(\L^{-1})^\mu_\nu}}
\def\lnm{{(\L^{-1})_\mu^\nu}}

\def\ul{u^{-1}}
\def\Sl{S^{-1}}
\def\Ul{U^{-1}}

\def\au{\a^\mu}
\def\ad{\a_\mu}
\def\an{\a^\nu}
\def\bd{b_\mu}
\def\gu{\g^\mu}
\def\gd{\g_\mu}
\def\ga{\g^\a}
\def\gb{\g^\b}
\def\gl{\g^\l}
\def\gn{\g^\nu}
\def\gr{\g^\rho}
\def\gs{\g^\s}
\def\ku{k^\mu}
\def\kd{k_\mu}
\def\kn{k^\nu}
\def\kl{k^\l}
\def\jb{j^\b}
\def\ju{j^\mu}
\def\jd{j_\mu}
\def\jn{j^\nu}
\def\pb{\p^\b}
\def\pu{\p^\mu}
\def\pd{\p_\mu}
\def\pl{\p^\l}
\def\pn{\p^\nu}
\def\pr{\p^\rho}
\def\ps{\p^\s}
\def\px{\p^x}
\def\qd{q_\mu}
\def\su{\s^\mu}
\def\sd{\s_\mu}
\def\xu{x^\mu}
\def\xd{x_\mu}
\def\xa{x^\a}
\def\xb{x^\b}
\def\xl{x^\l}
\def\xm{x'^\mu}
\def\xn{x^\nu}
\def\xs{x^\s}
\def\Au{A^\mu}
\def\Ad{A_\mu}
\def\An{A^\nu}
\def\Bu{B^\mu}
\def\Bd{B_\mu}
\def\Ju{J^\mu}
\def\Jd{J_\mu}

\def\qa{{\sss (\a)}}
\def\Qi{{\sss (i)}}
\def\qj{{\sss (j)}}
\def\qk{{\sss (k)}}
\def\ql{{\sss (\l)}}
\def\qlp{{\sss (\l')}}
\def\qn{\sss {(n)}}
\def\qp{{\sss (+)}}
\def\qm{{\sss (-)}}
\def\qH{{\sss (H)}}
\def\qpm{{\sss (\pm)}}
\def\qmp{{\sss (\mp)}}

\def\xbm{\f{\p\xb}{\p\xm}}
\def\xma{\f{\p\xm}{\p\xa}}
\def\xmn{\f{\p\xm}{\p\xn}}
\def\xnm{\f{\p\xn}{\p\xm}}
\def\xml{\f{\p\xm}{\p\xl}}
\def\xlm{\f{\p\xl}{\p\xm}}
\def\xns{\f{\p x'^\nu}{\p\xs}}
\def\xsn{\f{\p\xs}{\p x'^\nu}}
\def\fht{\f{\p\phi}{\p t}}
\def\fpt{\f{\p\psi}{\p t}}
\def\frt{\f{\p\rho}{\p t}}

\def\fAt{\f{\p\vA}{\p t}}
\def\fBt{\f{\p\vB}{\p t}}
\def\fEt{\f{\p\vE}{\p t}}

\def\dt{\f{d}{d t}}

\def\pt{\f{\p}{\p t}}
\def\ptt{\f{\p^2}{\p t^2}}
\def\pxt{\f{\p^2}{\p x^2}}
\def\pxm{\f{\p}{\p\xu}}
\def\pxn{\f{\p}{\p\xn}}
\def\pxdm{\f{\p}{\p\xm}}
\def\pxdn{\f{\p}{\p x'^nu}}

\def\rp{\not\!\p}
\def\rk{\not\! k}
\def\rA{\not\!\! A}
\def\rD{\not\!\!\cD}
\def\lrp{\stackrel{\leftarrow}{\p}}
\def\rrp{\stackrel{\rightarrow}{\p}}

\def\bk{{\bar{k}}}
\def\bp{{\bar{\psi}}}
\def\bq{{\bar{q}}}
\def\bu{{\bar{u}}}
\def\bv{{\bar{v}}}
\def\bx{{\bar{x}}}
\def\bC{{\bar{C}}}
\def\bH{{\bar{H}}}
\def\bV{{\bar{V}}}
\def\tp{{\dot{\psi}}}
\def\tq{{\dot{q}}}
\def\tqd{{\dot{\qd}}}
\def\tpg{{\dot{\psi}^\dagger}}
\def\tpp{{\dot{\psi}^{(+)}}}
\def\tpm{{\dot{\psi}^{(-)}}}
\def\dq{{\dot{q}}}
\def\dx{{\dot{x}}}
\def\dA{{\dot{A}}}
\def\ha{{\hat{a}}}
\def\hc{{\hat{c}}}
\def\hi{{\hat{\imath}}}
\def\hj{{\hat{\jmath}}}
\def\hk{{\hat{k}}}
\def\hn{{\hat{n}}}
\def\hp{{\hat{\psi}}}
\def\hr{{\hat{r}}}
\def\hs{{\hat{\s}}}
\def\hx{{\hat{x}}}
\def\hy{{\hat{y}}}
\def\hz{{\hat{z}}}
\def\hA{{\hat{A}}}
\def\hH{{\hat{H}}}
\def\hJ{{\hat{J}}}
\def\hN{{\hat{N}}}
\def\hS{{\hat{S}}}
\def\hvp{{\hat{\vp}}}
\def\hpu{{\hat{p}^\mu}}
\def\hpd{{\hat{p}_\mu}}
\def\hpp{{\hat{\psp}^\qp}}
\def\hpm{{\hat{\psp}^\qm}}
\def\hppg{{\hat{\psp}^{\qp\dagger}}}
\def\hpmg{{\hat{\psp}^{\qm\dagger}}}
\def\ta{{\tilde{a}}}
\def\td{{\tilde{d}}}
\def\tg{{\tilde{g}}}
\def\tn{{\tilde{\eta}}}
\def\tv{{\tilde{v}}}
\def\tx{{\tilde{\xi}}}
\def\tA{{\tilde{A}}}
\def\tD{{\tilde{D}}}
\def\tF{{\tilde{F}}}
\def\tG{{\tilde{G}}}
\def\tJ{{\tilde{J}}}

\def\va{{\mbox{\boldmath $\a$}}}
\def\vb{{\mbox{\boldmath $b$}}}
\def\vc{{\mbox{\boldmath $c$}}}
\def\ve{{\mbox{\boldmath $\e$}}}
\def\vg{{\mbox{\boldmath $\g$}}}
\def\vj{{\mbox{\boldmath $\jmath$}}}
\def\vm{{\mbox{\boldmath $m$}}}
\def\vnd{{\mbox{\boldmath $\n$}_\mu}}
\def\vo{{\bf 0}}
\def\vp{{\mbox{\boldmath $p$}}}
\def\vq{{\mbox{\boldmath $q$}}}
\def\vr{{\mbox{\boldmath $r$}}}
\def\vs{{\mbox{\boldmath $\s$}}}
\def\vt{{\mbox{\boldmath $\tau$}}}
\def\vv{{\mbox{\boldmath $v$}}}
\def\vx{{\mbox{\boldmath $x$}}}
\def\vy{{\mbox{\boldmath $y$}}}
\def\vz{{\mbox{\boldmath $z$}}}
\def\vA{{\mbox{\boldmath $A$}}}
\def\vB{{\mbox{\boldmath $B$}}}
\def\vBd{{\mbox{\boldmath $B$}_\mu}}
\def\vD{{\mbox{\boldmath $D$}}}
\def\vE{{\mbox{\boldmath $E$}}}
\def\vcE{{\mbox{\boldmath $\cE$}}}
\def\vF{{\mbox{\boldmath $F$}}}
\def\vH{{\mbox{\boldmath $H$}}}
\def\vJ{{\mbox{\boldmath $J$}}}
\def\vM{{\mbox{\boldmath $M$}}}
\def\vP{{\mbox{\boldmath $P$}}}
\def\vR{{\mbox{\boldmath $R$}}}
\def\vS{{\mbox{\boldmath $S$}}}
\def\vT{{\mbox{\boldmath $T$}}}
\def\vV{{\mbox{\boldmath $V$}}}
\def\vbd{\vb_\mu}
\def\curl{\mbox{\boldmath $\vn\!\times\!$}}
\def\div{\mbox{\boldmath $\vn\cdot$}}

\def\ma{m^\ast}
\def\ua{u^\ast}
\def\na{\eta^\ast}
\def\na{\eta^\ast}
\def\pa{\psi^\ast}
\def\xa{\xi^\ast}
\def\za{z^\ast}
\def\Ja{J^\ast}

\def\dag{\dagger}
\def\ag{a^\dagger}
\def\bg{b^\dagger}
\def\cg{c^\dagger}
\def\dg{d^\dagger}
\def\eg{e^\dagger}
\def\ig{\pi^\dagger}
\def\pg{\psi^\dagger}
\def\sg{\s^\dagger}
\def\ug{u^\dagger}
\def\Ag{A^\dagger}
\def\Bg{B^\dagger}
\def\Cg{C^\dagger}
\def\Sg{S^\dagger}
\def\Ug{U^\dagger}
\def\Vg{V^\dagger}
\def\hpg{\hp^\dagger}

\def\ppm{\psi^\qpm}
\def\psp{\psi^\qp}
\def\psm{\psi^\qm}
\def\pspg{{\psp}^\dagger}
\def\psmg{{\psm}^\dagger}
\def\upm{u^\qpm}
\def\up{u^\qp}
\def\um{u^\qm}
\def\Ab{A^\b}
\def\Al{A^\l}
\def\As{A^\s}
\def\Dd{D_\mu}
\def\Np{N^\qp}
\def\Nm{N^\qm}

\def\ppn{{\pa\n\psi-\psi\n\pa}}
\def\ppt{{\pa\pt\psi-\psi\pt\pa}}
\def\pfpt{{\pa\fpt-\psi\f{\p\pa}{\p t}}}
\def\ppu{{\pa\pu\psi-\psi\pu\pa}}
\def\ppvn{{\pa\vn\psi-\psi\vn\pa}}

\def\tm{\f{p^2}{2m}}
\def\hM{\f{\h^2}{2M}}
\def\hm{\f{\h^2}{2m}}
\def\ib{\f{i}{2}}
\def\iq{\f{i}{4}}
\def\ih{\f{i}{\h}}
\def\ihc{\f{i\h}{c}}
\def\ihm{\f{i\h}{2m}}
\def\hb{\f{\h}{2}}
\def\ph{\f{1}{2\pi}}
\def\tph{\f{2\pi}{\h}}
\def\mh{\f{1}{2m}}
\def\hf{\f{1}{2}}
\def\qt{\f{1}{4}}
\def\hg{\f{1}{2i}}
\def\Nf{\f{1}{N!}}

\def\mih{\left(-\ih\right)}

\def\bph{{(2\pi)^{1/2}}}
\def\bpb{{(2\pi)^2}}
\def\bpt{{(2\pi)^3}}
\def\bpf{{(2\pi)^4}}
\def\bpm{{(2\pi)^{-1}}}

\def\lg{\langle}
\def\rg{\rangle}
\def\la{\leftarrow}
\def\ra{\rightarrow}
\def\Ra{\Rightarrow}
\def\da{\downarrow}
\def\lra{\leftrightarrow}
\def\Ua{\uparrow}

\def\act{{\a\cdot\tau}}
\def\gck{{\g\cdot k}}
\def\kcx{{\vk\cdot\vx}}
\def\kcr{{\vk\cdot\vr}}
\def\kcR{{\vk\cdot\vR}}
\def\pcx{{\vp\cdot\vx}}
\def\qcr{{\vq\cdot\vr}}
\def\qcx{{\vq\cdot\vx}}
\def\scp{{\vs\cdot\vp}}
\def\pcA{{\p\cdot A}}
\def\ncA{{\n\cdot\vA}}
\def\ncB{{\n\cdot\vB}}
\def\ncE{{\n\cdot\vE}}
\def\tca{{\vt\cdot\va}}
\def\btB{{\b\t\vB}}
\def\ntA{{\n\t\vA}}
\def\ntB{{\n\t\vB}}
\def\ntE{{\n\t\vE}}

\def\anb{{\a\neq\b}}
\def\bna{{\b\neq\a}}
\def\inj{{i\neq j}}
\def\jni{{j\neq i}}
\def\mnn{{\mu\neq\nu}}

\def\skl{\sum_{\vk,\l}}
\def\sa{\sum_\a}
\def\sb{\sum_\b}
\def\sk{\sum_k}
\def\svk{\sum_{\vk}}
\def\svp{\sum_{\vp}}
\def\sm{\sqrt{\frac{m}{k_0}}}

\def\i{\infty}
\def\iii{\int_{-\infty}^\infty}
\def\iit{\int_{-\infty}^t}
\def\iti{\int_t^\infty}
\def\itt{\int_{t_0}^t}
\def\ini{\int_0^\infty}
\def\iin{\int_{-\infty}^0}
\def\inb{\int_0^\b}
\def\inp{\int_0^\pi}

\def\fc{{\bf\chi}}
\def\fe{{\bf\e}}
\def\fp{{\bf\phi}}
\def\fv{{\bf\varphi}}
\def\fA{{\bf A}}
\def\fB{{\bf B}}
\def\fD{{\bf D}}
\def\fL{{\bf L}}
\def\fF{{\bf F}}
\def\fH{{\bf H}}
\def\fK{{\bf K}}
\def\fV{{\bf V}}

\def\bfH{\bar{\fH}}
\def\bfV{\bar{\fV}}
\def\vfA{\mbox{\boldmath $\fA$}}

\def\cD{{\cal D}}
\def\cE{{\cal E}}
\def\cF{{\cal F}}
\def\cH{{\cal H}}
\def\cL{{\cal L}}
\def\cX{{\cal X}}
\def\cDu{\cD^\mu}
\def\cDd{\cD_\mu}

\def\vu{{\bm{c}1\\0\em}}
\def\vd{{\bm{c}0\\1\em}}
\def\vam{{\bm{cc}0&\vs\\\vs&0\em}}
\def\vbm{{\bm{cc}\vs&0\\0&\vs\em}}
\def\II{{\bm{cc}I&0\\0&I\em}}
\def\Is{{\bm{cc}1&0\\0&1\em}}
\def\It{{\bm{cccc}1&0&0&0\\0&1&0&0\\0&0&1&0\\0&0&0&1\em}}
\def\I0{{\bm{cc}1&0\\0&1\em}}
\def\Ia{{\bm{cc}0&I\\I&0\em}}
\def\Ib{{\bm{cc}0&-I\\I&0\em}}
\def\Ic{{\bm{cc}I&0\\0&-I\em}}
\def\sxm{{\bm{cc}0&1\\1&0\em}}
\def\sym{{\bm{cc}0&-i\\i&0\em}}
\def\szm{{\bm{cc}1&0\\0&-1\em}}
\def\Sxm{{\bm{cc}0&\s_x\\-\s_x&0\em}}
\def\Sym{{\bm{cc}0&\s_y\\-\s_y&0\em}}
\def\Szm{{\bm{cc}0&\s_z\\-\s_z&0\em}}
\def\gm{{\bm{cccc}1&0&0&0\\0&-1&0&0\\0&0&-1&0\\0&0&0&-1\em}}

\def\le{\langle}
\def\re{\rangle}

\def\Akl{\vA_\vk^\ql}
\def\Akla{\vA_\vk^{\ql\ast}}
\def\aql{a^\ql}
\def\aqld{{\aql}^\dagger}
\def\qql{q^\ql}
\def\qqld{{\qql}^\dagger}
\def\vvl{\vcE^\ql}
\def\vvla{\vcE^{\ql\ast}}
\def\vml{\cE_\mu^\ql}

\def\dtk{d^3\!k}
\def\dtx{d^3\!x}
\def\dtp{d^3\!p}
\def\dfk{d^4\!k}
\def\dfx{d^4\!x}
\def\dfp{d^4\!p}

\def\tf{\dot{.~.}\ }
\def\ie{\stackrel{\textstyle .}{..}}
\def\due{\raisebox{-.8mm}{$\stackrel{\textstyle .~.}{\stackrel{}{.}}$}}
\def\LQED{\cL_{\scriptscriptstyle QED}}
\def\LEM{\cL_{\scriptscriptstyle EM}}
\def\LYM{\cL_{\scriptscriptstyle YM}}
\def\LSO{\cL_{\scriptscriptstyle S.O.}}

\begin{center}
{\Large \bf Undulatory Variation of Antiferromagnetic Strength with Magnetic Field Based on Hubbard Model Hamiltonian}\\
\vskip1cm

Hyeonjin Doh and Sung-Ho Suck Salk\\
Department of Physics, Pohang University of Science and Technology\\
Pohang, Kyungbook 790-784, Korea\\
\vskip2cm

{\bf abstract}
\end{center}
\hspace*{0.6cm}Using the Hubbard model Hamiltonian in
a mean field level,
we examine the variation of antiferromagnetic strength
with applied magnetic field. It is demonstrated that minima 
in the
antiferromagnetic strength exist at the the even integer denominator values of
rational number for magnetic flux per plaquette.
The undulatory behavior of antiferromagnetic strength 
with the external magnetic field is found. It is seen to be related 
to the undulatory net statistical phase 
owing to the influence of the applied magnetic field.
\newpage
\bi
\item[I.]{\large\bf Introduction}\\
\hspace*{0.6cm}Earlier Hasegawa et.al.[1] examined the total kinetic energy of spinless non-interacting electrons in  two-dimensional lattices as a function of magnetic flux per plaquette and found an absolute minimum for the case of one flux quantum per particle for a square lattice. 
One of the most widely used computational methods for the
studies of interacting electron systems is 
the Hubbard model Hamiltonian [2] 
for  various investigation of physical properties [3-5].
In the present study, we pay attention to the variation of antiferromagnetic strength, band gap and total energy  with the applied magnetic field for the system of interacting electrons in a two-dimensional square lattice.
\item[II.]{\large\bf Variation of Antiferromagnetic Strength, Total Energy and Band Gap with External Magnetic Field}\\
\hspace*{0.6cm}With the external magnetic field, the one-band Hubbard Hamiltonian is written,
\bn H = -t \ds{\sum_{\lg i, j \rg \s}} \exp [(-i2\pi/\phi_0) \int^{i}_{j} Adl] C^{+}_{i \s} C_{j \s}
+ U \ds{\sum_{j}} n_{j \Ua} n_{j \da}-\mu \ds{\sum_{j}} (n_{j \Ua} + n_{j \da}) \en
Here $t$ and $U$ are the hopping integral and the Coulomb repulsion energy respectively. $\mu$ is the chemical potential.
$\lg i, j\rg$ stands for summation only over the nearest neighbors. $n_{j \s}$ is the number operator for an electron of spin $\s$ at site $j$, 
$n_{j\s} = C^{+}_{j \s} C_{j \s}$ with $C^{+}_{j\s} (C_{j\s})$, the creation (annihilation) operator for an electron at site $j$.
$\phi_0$ is the magnetic flux quantum, $\phi_0 = \f{hc}{e}$.
The electromagnetic vector potential in the Landau gauge is $A=B(0,x,0)$ with $x=ma$ for the m-th site along y direction with lattice spacing a. 
Thus $\int A dl = m Ba^{2} = m\phi$ with $\phi = Ba^2$. 

\hspace*{0.6cm}Applying the mean field (Hartree-Fock) approximation [6] to the Hubbard model Hamiltonian 
with the use of the two component spinor, 
\bn \psi_j  = \left( \ba{c} C_{j \Ua}\\ C_{j \da} \ea \right),\en
and the spin operators,
\bn S_z (j) = \f{1}{2} (C^{+}_{j \Ua}  C_{j \Ua} - C^{+}_{j \da} C_{j \da}), \en
and 
\bn S_{+} (j) = C^{+}_{j \Ua} C_{j \da},~S_{-}(j) = C^{+}_{j \da} C_{j \Ua}, \en
one readily obtains
\bn\ba{c}
H = -t\ds{\sum_{\lg i j\rg}}\exp(-i 2\pi m \phi/\phi_0)\psi^{+}_{i} \psi_{j}+U \ds{\sum_{j}} \psi^{+}_{j} [\f{1}{2} \lg n(j)\rg - \lg S(j)\rg]\psi_j\\
+ U \ds{\sum_j} \{\lg S_z (j)\rg^2 + \lg S_{+}(j)\rg\lg S_{-} (j) \rg -\f{1}{4} \lg n(j)\rg^2 \}\\
- \mu \ds{\sum_{j}} (n_{j\Ua} + n_{j\da}),\ea\en
where \bn n(j) = C^{+}_{j\Ua} C_{j\Ua} + C^{+}_{j\da} C_{j\da}\en
and
 \bn \lg S(j) \rg  = \left( \ba {cc}S_z (j)& S_{-}(j) \\
S_{+}(j)& -S_{z} (j)\ea\right), \en
where $S_z (j), S_{+}(j)$ and $ S_{-}(j) $  denote the average values of $S_z, S_+$ and $S_{-}$ respectively in the ground state.

\hspace*{0.6cm}For non-interacting spinless electrons, i.e., $U=0$,
we performed numerical calculations of the total kinetic energies per site at various values of both electron filling factor, $\nu$ and magnetic flux per plaquette, $\phi = (p/q)\phi_0$ 
for the finite size square lattice of $24\times 24$. Our calculations[7] yielded excellent agreements with Hasegawa et. al.$'$s exact results[1] for the square lattice of infinite size. 
In Table 1, for the ($24\times 24$) antiferromagnetic square lattice of a half-field band we show the variation of the total electronic energy with the flux quantum per plaquette at various values of Coulomb repulsion energy. 
For the non-interacting electrons, the total kinetic energies agreed extremely well with the exact results of Hasegawa et al.
It is of note that the tabulated total energy at $\phi = \f{1}{2}\phi_0$ with $U=0$ is twice the value for the case of the integer quantum Hall effect(IQHE) with quantum number 1.
For the present case of the half-filled band it corresponds to the IQHE with quantum number 2. It is of note that the absolute minima of the total energies are always found at the flux value of $\phi= \f{1}{2} \phi_0$, regardless of whether electron correlation is present ($U\neq 0$) or not $(U=0)$.
In the presence of a periodic potential the total electronic ground state energy in an external magnetic field is found to be always smaller than the case of zero magnetic field.
It is also of note that the total energy difference between zero and non-zero magnetic fields decreases with the strength of electron correlation. 

\hspace*{0.6cm}We define $\a = \ds{\sum_i} | \lg S_z (i)\rg |^2 /N$ as antiferromagnetic `strength' or loosely `order' parameter to estimate the variation of antiferromagnetic strength with the applied magnetic field. 
In Fig. 1 we show the variation of $\a$ as a function of the magnetic flux $\phi = (p/q)\phi_0$ for various selected values of $U$. 
The solid lines indicate the computed results for the $20 \times 20$ square lattice.
Symbols other than the solid line represent various sizes of square lattices up to the size of $36 \times 36$ with which the periodic boundary conditions are satisfied.
Despite the failure of meeting the periodic boundary condition with q = 3, 5, 8 in the rational number of $p/q$, it was found that the $20 \t 20$ square lattice yielded reasonably good agreements with the results from other square lattices which satisfy the periodic boundary conditions [7]. 
The variation of antiferromagnetic strength (order parameter) $\a$ with the magnetic flux $\phi$ diminishes at substantially high values of U,
e.g., $U=20t$.
It is to be noted that there exists a strong tendency of losing the 
antiferromagnetic strength at $\phi=\f{1}{2} \phi_0$ even with relatively strong correlation, e.g., $U=3t$ as shown in Fig.1. 
Unless correlation (Coulomb repulsion) is exceedingly strong, the absolute minima are found to occur at $\phi=\f{1}{2} \phi_0$.
With the lower values of $U$, there exists undulatory variation of the antiferromagnetic strength $\a$ with the magnetic field.
Interestingly, local minima are found only at the even integer values of q (denominator) for the magnetic flux per plaquette, 
particularly so at larger values of $\phi = (p/q) \phi_0$,  e.g., $\f{1}{2}, \f{3}{8}, \f{1}{4},
\frac{1}{6}$,  and $\f{1}{8}$, while at the odd integer values of q such local minima do not exist.
To explore the cause of such differences in antiferromagnetic strength between the even and odd integer values of $q$, we computed the band gap as a function of magnetic field.
As shown in Fig.2, the band gap also showed local minima at the even integer values of q. Consequently,
easier electron hopping at the even integer values of the denominator for the flux tends to destroy the antiferromagnetic strength.
A noticeable band gap opening begins to occur above $U=2t$.
As shown in both Fig.2 and Fig.3, for all values of Coulomb repulsion energy $U$ the absolute minimum of band gap is also found to occur invariably at $\phi = \f{1}{2} \phi_0$, thus causing absolute minima in antiferromagnetic strength at this value of magnetic flux.

\hspace*{0.6cm}The generalized equal time commutation relation for anyons with hard cores on a square lattice is given by the Jordan-Wigner transformation[3] 
\be a_i a^{+}_{j} = \d_{ij} - e^{i\d} a^{+}_{j} a_i \ee
where $a^{+}_{i}$ and $a_j$ are the anyon creation and annihilation operators on the sites $i$ and $j$ of the square lattice.
The angle $\d$ represents a phase associated with the statistical (Chern-Simons) gauge field.
The Heisenberg antiferromagnet on a square lattice is equivalent to an arrangement of spinless fermions coupled to
a Chern-Simons gauge field. 
The present study deals with the role of the 
external magnetic field on the square
lattice of antiferromagnet.
Our computed results showed the absolute minima of the antiferromagnetic strength $\a$ with the magnetic flux of $\f{1}{2}\phi_0$ per plaquette.
This indicates that part of the statistical (Chern-Simons) gauge field coupled to the spinless electrons is cancelled out by the contribution of the electromagnetic vector potential (gauge field).
For the system of relatively weak electron correlation, say, $U\leq 2t$, the effective (net) statistical gauge field is found to be small or negligible for the spinless fermions at the value of $\phi = \f{1}{2} \phi_0$. 
The undulatory behavior of the antiferromagnetic strength discussed earlier is now noted to be related to the undulation of the effective (net)  statistical phase (or net statistical transmutation) of spinless fermions coupled to the Chern-Simons
gauge field which is partially cancelled out by contributions from the external magnetic field. 
In summary the undulatory cancellation of the statistical (Chern-Simons) gauge field by the electromagnetic gauge field is maximized at the value of the magnetic
flux, $\phi = \f{1}{2} \phi_0$. Thus this results in the spinless fermions coupled to the substantially reduced net gauge field for the case of relatively weak correlation.
\newpage
\item[III.]{\large \bf Conclusion}\\
\hspace*{0.6cm}In the present study we examined the variations of the total electronic ground state energy, band gap, and antiferromagnetic strength(order) with the magnetic flux for various strengths of electron correlation. 
The absolute minima of antiferromagnetic strength, total electronic energy and band gap were found to invariably occur at the value of magnetic flux, $\phi= \f{1}{2}\phi_0$ per plaquette. 
For the case of the total energy the absolute minimum at $\phi = \f{1}{2} \phi_0$ was shown to be universal, regardless of whether electron correlation is present or not.
For $U \neq 0$, the undulatory antiferromagnetic strength $\a$ was found to give local minima at the even integer values of denominators for the magnetic flux per plaquette, that is, $\f{1}{2}, \f{3}{8}, \f{1}{4}$ and $\f{1}{8}$.
Unless electron correlation is sufficiently strong, the undulatory variation of antiferromagnetic strength with magnetic field is found to be related to the undulatory variation of the effective statistical phase, which results from coupling of the spinless fermions to the statistical (Chern-Simons) gauge field that is affected (partially cancelled) by the electromagnetic field. 
At the even integer denominator values of the magnetic flux, electrons 
are seen to behave like spinless non-interacting fermions due to coupling to a small or negligible net (effective) gauge field particularly with relatively weak correlation, say, $U \leq 2t$, thus showing equivalence to the integer quantum Hall effect.
\ei
\newpage
{\bf Acknowledgement :}\\
One(SHSS) of us is supported by the Korean Ministry of Education BSRI program and by the Center for Molecular Science at Korea Advanced Institute of Science and Technology.  We are grateful to professor H. Y. Choi for helpful discussions.
\bi
\item[Table 1 :] 
Variation of the total electronic energy per site with 
 magnetic flux per plaquette, $\phi = (p/q) \phi_0$ for the $(24 \times 24)$ 
antiferromagnetic square lattice of a half-field band at various values of 
Coulomb repulsion energy $U$. 

\item[Fig. 1 :] Change of antiferromagnetic strength (order) parameter $\a$ with magnetic flux per plaquette, $\phi$ at various strengths of electron correlation, $U$.\\
\item[Fig. 2 :]Variation of band gap with magnetic flux per plaquette, $\phi$ at various strengths of electron correlation, $U$.\\
\item[Fig. 3 :]Band gap vs. Coulomb repulsion energy, $U$ as a function of magnetic flux per plaquette, $\phi$.\\
\ei
\newpage
{\bf Reference}
\bi
\item[[1]]J. Hubbard, Proc. R. Soc. London Ser. A 276(1963) 238; 281 (1964) 401
\item[[2]]The Hubbard Model : A Reprint Volume, ed. A. Montorosi (World Scientific, Singapore, 1992)
\item[[3]]E. Fradkin, Field Theories of Condensed Matter Systems (Addison-Wesley, New York, 1991) ; references therein
\item[[4]]N. F. Mott, Metal Insulator Transitions (Taylor and Francis, London, 1974) ; references therein
\item[[5]]Y. Hasegawa, P. Lederer, T. M. Rice, and P. B. Wiegmann, Phys. Rev. Lett. 63(1989) 907
\item[[6]]H. Y. Choi, Phys. Rev. B 44 (1991)2609
\item[[7]]H.Doh and Sung-Ho Suck Salk, Physica C, in press;
          H.Doh and Sung-Ho Suck Salk, J. Kor. Phys. Soc. 28(1995) S588
\ei
\newpage
Table 1: \\
\begin{tabular}{lcccccc} 
 $U\backslash\phi$ & $0$ & $\frac{1}{8}$ & $\frac{1}{4}$ 
 &$\frac{1}{3}$&$\frac{3}{8}$&$\frac{1}{2}$ \\ \hline 
$0.0$&$-1.617$&$-1.650$&$-1.717$&$-1.712$&$-1.758$&$-1.915$ \\ 
$0.5$&$-1.743$&$-1.775$&$-1.842$&$-1.838$&$-1.883$&$-2.040$ \\ 
$1.0$&$-1.870$&$-1.900$&$-1.967$&$-1.965$&$-2.009$&$-2.165$ \\ 
$1.5$&$-2.001$&$-2.027$&$-2.092$&$-2.097$&$-2.139$&$-2.290$ \\ 
$2.0$&$-2.139$&$-2.160$&$-2.218$&$-2.235$&$-2.275$&$-2.415$ \\ 
$3.0$&$-2.447$&$-2.462$&$-2.500$&$-2.533$&$-2.565$&$-2.666$ \\ 
$4.0$&$-2.797$&$-2.808$&$-2.836$&$-2.865$&$-2.886$&$-2.935$ \\ 
$20.0$&$-10.198$&$-10.198$&$-10.198$&$-10.199$&$-10.199$&$-10.200$ \\ 
\end{tabular}

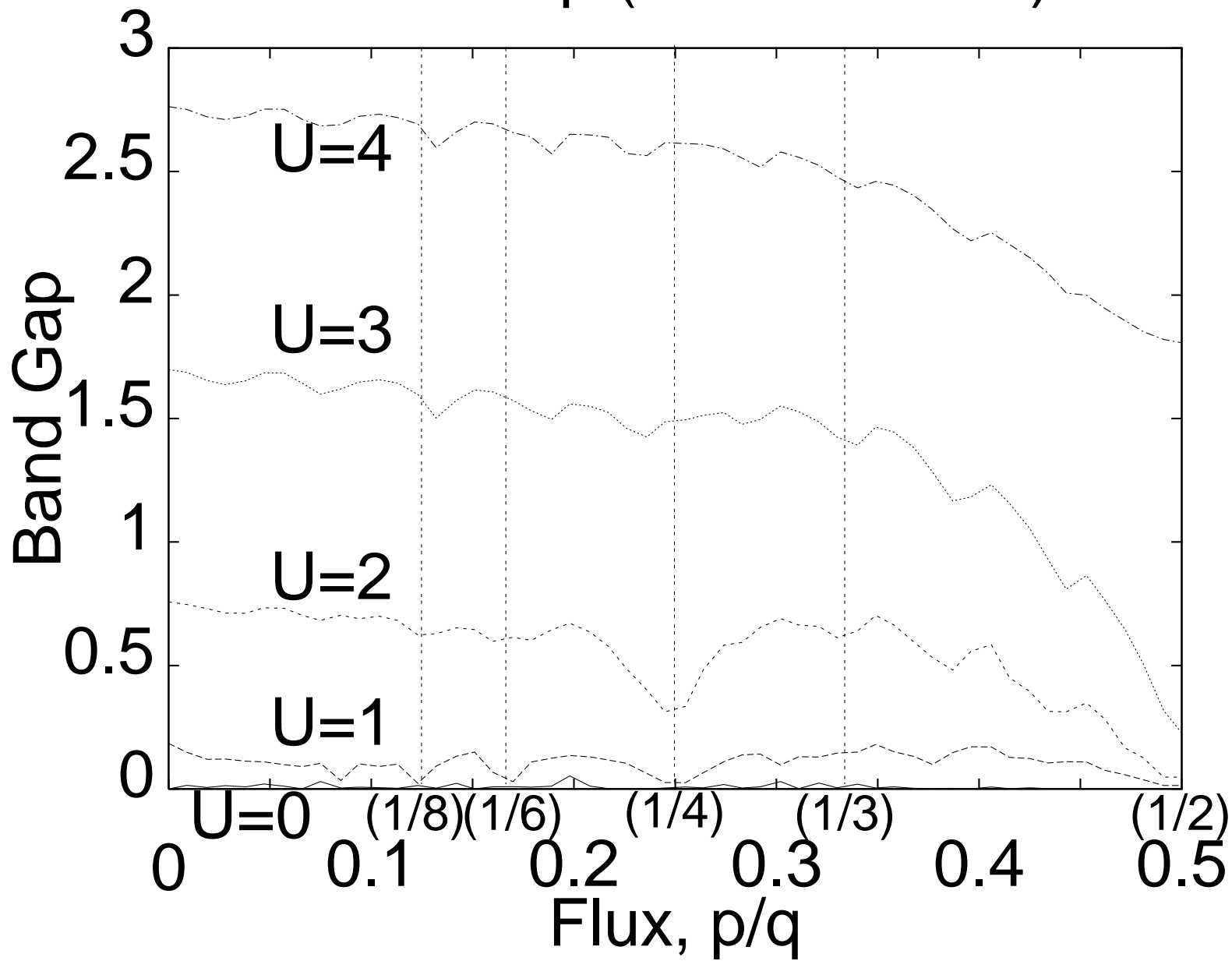
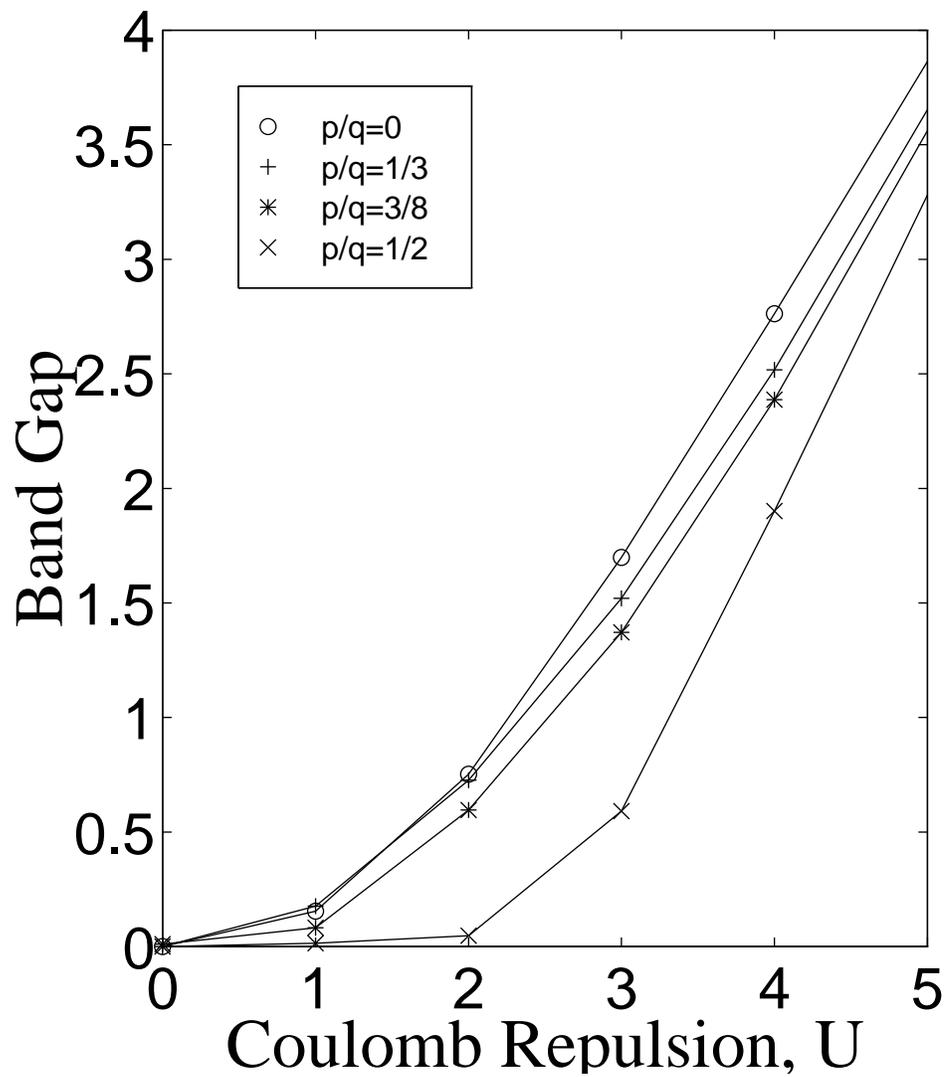
\end{document}